\documentclass[fleqn,twoside,twocolumn,nofootinbib,showkeys]{revtex4}
\usepackage[sec,nocpr]{ujp}
\usepackage{amsmath}
\usepackage{dsfont}
\usepackage{color}
\usepackage{verbatim}

\usepackage[pdfstartview={FitH top},pdfpagemode=None,breaklinks]{hyperref}
\hypersetup{
    colorlinks,
    citecolor=black,
    filecolor=black,
    linkcolor=black,
    urlcolor=black
}

\allowdisplaybreaks
\begin{document}
\title[Deformed Bose Gas Models]
{DEFORMED BOSE GAS MODELS AIMED\\ AT TAKING INTO   ACCOUNT BOTH
COMPOSITENESS\\
OF PARTICLES AND THEIR INTERACTION$^1$} 

\author{A.M. Gavrilik}
\affiliation{Bogolyubov Institute for Theoretical Physics, Nat. Acad. of Sci. of Ukraine}
\address{14b, Metrolohichna Str., Kyiv 03680, Ukraine}
\email{omgavr@bitp.kiev.ua}

\author{Yu.A. Mishchenko}
\affiliation{Bogolyubov Institute for Theoretical Physics, Nat. Acad. of Sci. of Ukraine}
\address{14b, Metrolohichna Str., Kyiv 03680, Ukraine}

\udk{530.145, 536.71,\\[-3pt] 539.1} \pacs{02.20.Uw, 05.30.-d,\\[-3pt] 05.30.Jp, 05.90.+m,\\[-3pt] 05.70.Ce, 11.10.Lm,\\[-3pt] 46.25.Cc} \razd{\secix}

\autorcol{A.M.\hspace*{0.7mm}Gavrilik, Yu.A.\hspace*{0.7mm}Mishchenko}%

\setcounter{page}{1171}%

\begin{abstract}
We consider the deformed Bose gas model with the deformation
structure function that is the combination of a $q$-deformation and
a quadratically polynomial deformation. Such a choice of the
unifying deformation structure function enables us to describe the
interacting gas of composite (two-fermionic or two-bosonic) bosons.
Using the relevant generalization of the Jackson derivative, we
derive a two-parametric expression for the total number of
particles, from which the deformed virial expansion of the equation
of state is obtained. The latter is interpreted as the virial
expansion for the effective description of a gas of interacting
composite bosons with some interaction potential.
\end{abstract}


\maketitle

\section{Introduction}

The treatment of real non-ideal gases in statistical physics or
thermodynamics, which involves the interaction and the composite
nature of particles, deals usually with special techniques or
approximations. To take the interaction between particles into
account, the virial expansions~\cite{Landau_5,Huang,Pathria} of
thermodynamic relations, the equation of state, {\it etc.}  are
analyzed at small densities. There also exist several approaches
that account for the compositeness of particles and work in the
second quantization scheme with exact transformations and results
(i.e., without any simplifications), see, e.g.,
\cite{Girardeau1975,Hadjimichef1998,Combescot_ManyBody}. However,
from the practical point of view, the direct account of the
compositeness along with interaction between particles is quite
complicated and hardly realizable.

The concept of deformed oscillators
\cite{Meljanac,Man'ko1997,Bonatsos,Daskaloyannis_1991} and deformed
Bose-(or
Fermi-)gas~\cite{Lavagno_Thermostat,Lavagno_Therm,GR_Virial} is one
of the possibilities to consider the above-mentioned factors in an
effective way (approximately, as a rule). This is achieved by means
of the realization (see some details in~\cite{GKM2} and Section 2
below) of a gas of composite particles or a gas of pointlike
interacting particles in terms of a specially constructed model of
gas of deformed (quasi)particles. The systems of composite
two-component Bose-type particles (called quasibosons) can be
modeled by the corresponding systems (gases, {\it etc.}) of
$q$-deformed bosons, as explicitly shown in~\cite{Avan}. There are
also some other
works~\cite{Bagheri_Harouni,Liu,Zeng2011,Greenberg_Statist,Sviratcheva,Bonatsos_Mol}
on the effective description of composite particles (excitons,
nucleons, molecules, {\it etc.}) by the use of deformed ones.
Recently, the two-fermionic (and also two-bosonic) composite bosons
were considered~\cite{GKM,GKM2} from the viewpoint of their
algebraic realization by deformed bosons. Namely, it was shown that
the realization on the states is possible for a certain quadratic
version of the deformation. Another version of the $q$-deformed Bose
gas model was applied to the systems of interacting particles
in~\cite{Scarfone_Interact}.\footnotetext[1]{This work is the
contribution to Proceedings of the International Conference
``Quantum Groups and Quantum Integrable Systems''.} Therein, the
corresponding deformed virial expansion in powers of a deviation
$\epsilon=q-1$, where $q$ is the deformation parameter, was
analyzed. Moreover, for a $p,q$-deformed Bose gas, the virial
expansion was treated in~\cite{GR_Virial}, and some interesting
implications were drawn. In addition, the thermodynamics of diverse
deformed Bose- and Fermi-gases was studied in detail (see, e.g.,
\mbox{\cite{Lavagno_Therm,Algin2012}}).

Combining the ideas of~\cite{Scarfone_Interact} and [14, 22], we
propose a unified deformed model composed of the $q$-deformation and
the quadratic deformation \cite{GKM,GKM2} to effectively describe a
gas of interacting composite (two-fermionic or two-bosonic) bosons.
Here, the thermodynamical aspects, more specifically the virial
expansion of the equation of state, are under
study.\looseness=-1\vspace*{-1.5mm}

\section{Preliminaries}\label{sec:prelim}

Let us mention the earlier obtained results on the ef\-fec\-ti\-ve
description of a gas of composite bosons with interaction between
particles in terms of deformed oscillators, which serve as our
starting point.\looseness=-1\vspace*{-1.5mm}

\subsection{Compositeness aspects}

Composite bosons constructed of two fermions (or two bosons) in the
second quantization scheme have the creation and annihilation
operators in the form \mbox{[14,\,15,\,22]}\vspace*{-1.5mm}
\begin{equation}
A^\dag_\alpha = \sum_{\mu\nu} \Phi_\alpha^{\mu\nu} a^\dag_\mu b^\dag_\nu,\quad
A_\alpha = \sum_{\mu\nu} \overline{\Phi_\alpha^{\mu\nu}} b_\nu a_\mu,
\end{equation}\vspace*{-3mm}

\noindent  where $a^\dag_\mu$, $b^\dag_\nu,$ and $a_\mu$, $b_\nu$
are the creation and annihilation operators for the constituents.
The matrices $\Phi_\alpha^{\mu\nu}$ are related to the constituents'
wavefunctions. The commutator between $A_\alpha$ and $A^\dag_\beta$
is of special interest for calculations and the analysis, and it is
of the form\vspace*{-1.5mm}
\begin{align*}
&[A_\alpha,A^\dag_\beta] = \delta_{\alpha\beta} - \Delta_{\alpha\beta}, \\
&\Delta_{\alpha\beta} = \sum_{\mu\mu'}(\Phi_{\beta}{\Phi}^{\dag}_{\alpha})^{\mu'\mu}
a^{\dag}_{\mu'}a_{\mu} + \sum_{\nu\nu'}({\Phi}^{\dag}_{\alpha}
\Phi_{\beta})^{\nu\nu'} b^{\dag}_{\nu'}b_{\nu}.
\end{align*}\vspace*{-3mm}

\noindent Here, $\Delta_{\alpha\beta}$ gives a deviation from the
purely bosonic relation. If it is described merely in terms of
$A_\alpha$ and $A^\dag_\alpha,$ then we may speak about the
realization of composite bosons' system. Such a situation is very
useful from the viewpoint of simplification. The many-body system of
composite (two-fermionic) bosons with certain components' wave
functions can be realized, as shown in~\cite{GKM,GKM2}, by a
deformed Bose gas model with the quadratic structure function
$\phi_{\tilde{\mu}}(n) = (1+\tilde{\mu})n - \tilde{\mu} n^2$ of the
deformation and the discrete deformation parameter
$\tilde{\mu}=1/m$, $m=1,2,...$\,. According to the realization,
under certain conditions, the gas of composite bosons can be treated
on the states as the corresponding gas of deformed bosons. We remark
that, for composite bosons realized by deformed oscillators, the
characteristics of the intercomponent entanglement are remarkably
expressed~[25, 26] through the deformation
\mbox{parameter.}\vspace*{-1.5mm}

\subsection{Account of the interaction\\ between Bose particles}

The $q$-deformed algebra with the structure function $\phi_q(n) =
\frac{1-q^n}{1-q} \equiv [n]_q$ was used in~\cite{Scarfone_Interact}
to describe an interacting gas of  Bose particles. Therein, the
effects of the interaction between particles of a Bose gas were
incorporated, by using a deformed version of the model, and were
explicitly demonstrated with the use of $q$-deformed thermodynamic
relations. In particular, this concerns the expression for the
specific volume given in terms of the fugacity, as well as the
equation of state. We note that the deformed relations (e.g., for
the total number of particles) can be
obtained~\cite{Scarfone_Interact} by the use of the $q$-deformed or
Jackson derivative instead of the non-deformed one. A somewhat more
specific outline of the description of the systems of interacting
particles using the $q$-deformation is given in Section
\ref{sec:interact}.

In view of the already-mentioned fact of the realizability of
composite bosons, we utilize a deformed Bose gas with (the
quadratic) structure function $\phi_{\tilde{\mu}}(n)$ to find
effective thermodynamic relations or functions for the ideal
(non-interacting) {\it quantum gas of composite bosons}. Those
include the deformed (thus, depending on $\tilde{\mu}$) virial
expansion of the equation of state. To take the interaction between
particles simultaneously with their compositeness into account, the
two structure functions $\phi_{\tilde{\mu}}(n)$ and $\phi_q(n)$ can
be combined to give the unifying deformation structure function
$\phi_{\tilde{\mu},q}(n) = (1+\tilde{\mu})[n]_q - \tilde{\mu}
([n]_q)^2$. The latter will play a central role in our treatment. An
alternative version of the unifying structure function will be also
discussed.

In Sections 3 and 4, these two separately treated situations are
overviewed in more details.\vspace*{-1.5mm}

\section{Systems of Interacting Particles \\ Described by a \boldmath$q$-Deformed Algebra}\label{sec:interact}

Let us give a  brief overview of the results
of~\cite{Scarfone_Interact} concerning the interpretation of
interacting many-boson systems using $q$-deformed oscillators
($q$-bosons) without interaction. Each interpretation is based on
the assumption that a suitably chosen $q$-deformed thermodynamic or
statistical relation for non-interacting pointlike particles' system
can be applied (at least within  some  approximation) to an
interacting many-particle system with certain
\mbox{interaction.}\looseness=-1

In work~\cite{Scarfone_Interact}, a system of interacting Bose
particles described by the $q$-oscillator algebra is considered. As
a starting point, the series expansion of the basic number $[N]_q$
in terms of the ``deviation'' $\epsilon\equiv q-1$ is
taken:\vspace*{-2mm}
\begin{align}
&[N]_q \!=\! \Bigl(1\!-\!\frac{\epsilon}{2}\!+\!\frac{\epsilon^2}{3}\!-\!\frac{\epsilon^3}{4}\!+\!...\Bigr) N \!+\!
\frac{\epsilon}{2!} \Bigl(1\!-\!\epsilon\!+ \frac{11}{12}\epsilon^2\!-\!...\Bigr)\times\nonumber\\
&\times N^2 + \frac{\epsilon^2}{3!}
\Bigl(1\!-\!\frac32\epsilon\!+\!...\Bigr) N^3 +
O(N^4).\label{[N]-expansion}
\end{align}
This expansion can be naturally interpreted as the one incorporating
the interparticle interaction. According
to~\cite{Scarfone_Interact,Parthasarathy1992}, the contributions
from the interaction can be viewed either in terms of $N$, $N^2$,
$N^3$,~... or in terms of $\epsilon$, $\epsilon^2$, $\epsilon^3$,
...\,. In the latter case (for which the contributions from the
interaction are considered as those contained in the terms depending
on the deformation parameter), expansion~(\ref{[N]-expansion}) is
rewritten in the ``perturbative'' form\vspace*{-2mm}
\begin{align}
&[N]_q = \sum_{i=0}^\infty \frac{\epsilon^i}{(i+1)!} \prod_{j=0}^i (N-j) =\nonumber\\
&= N \!+\! \frac{\epsilon}{2!} N(N\!-\!1) \!+\! \frac{\epsilon^2}{3!} N(N\!-\!1)(N\!-\!2) \!+\! O(\epsilon^3).\label{[N]eps-exp}
\end{align}

Another relevant quantity needed to be considered is the Hamiltonian
$H_\epsilon = \frac12\omega ([N+1]_q+[N]_q)$ for a system of
$q$-bosons (single-mode case). The expansion of $H_\epsilon$ in
powers of $\epsilon$ linked with the deformation parameter $q$ is
given by~\cite{Scarfone_Interact}\vspace*{-2mm}
\begin{equation}\label{H_eps-exp}
H_\epsilon = H_0 + \omega \sum_{i=1}^\infty \epsilon^i \frac{(2N+1-i)}{2(i+1)!} \prod_{j=0}^{i-1}
(N-j).
\end{equation}
Similarly to (\ref{[N]eps-exp}), the terms of the first and higher
orders in $\epsilon$ in series~(\ref{H_eps-exp}) are again
interpreted as those arising from the interaction. To emphasize the
physical meaning, they constitute nothing but the interaction
Hamiltonian.

Remark that a direct relation of the deformation parameter $q$ to
the parameter(s) of an equivalent interaction Hamiltonian for
many-particle systems was not explicitly derived
in~\cite{Scarfone_Interact}, and that constitutes a nontrivial, but
important problem.

The idea of a $q$-deformed non-interacting (ideal) many-body system as a
non-deformed, but interacting system is best illustrated by means of the virial expansion, at least at this stage.
Basing on the deformed thermodynamic relations (particularly, the specific volume $v$ as a function of
the fugacity $z$ and the deformed equation of state) derived in~\cite{Lavagno_Thermostat,Lavagno_Therm}
for a gas of $q$-bosons, the  $q$-deformed virial expansion was obtained in the form
\begin{equation}
\frac{Pv}{k_{\rm  B} T} = \sum_{k=1}^\infty a_k(\epsilon)
\Bigl(\frac{\lambda^3}{v}\Bigr)^{k-1},
\end{equation}
where $a_k(\epsilon)$ denote the corresponding virial coefficients.
The first several coefficients given up to $\epsilon^3$
are~\cite{Scarfone_Interact}:
\begin{align*}
&a_1(\epsilon) = 1,\quad
a_2(\epsilon) = -\frac{1}{4\sqrt{2}}\!-\!\frac{1}{48\sqrt{2}} \epsilon^2(1\!-\!\epsilon) \!+\! O(\epsilon^4),\\
&a_3(\epsilon) = -\Bigl(\frac{2}{9\sqrt{3}}\!-\!\frac18\Bigr) \!-\! \Bigl(\frac{1}{18\sqrt{3}}\!-\!\frac{1}{48}\Bigr)
\epsilon^2(1\!-\!\epsilon) \!+\! O(\epsilon^4).
\end{align*}
Similarly to the previous cases, $\epsilon\ne0$ corrections to the
standard virial coefficients of the ideal quantum Bose gas can be
 interpreted as those arising from some explicitly
accounted interaction (certain interaction potential in the
Hamiltonian). Note that, on the other hand, this interaction can be
viewed in terms of $q$-bosons as such that arises from (the change
of) their quantum statistics and, thus, is of the quantum
statistical origin. This is in some analogy with the effective
interaction due to the Pauli exclusion principle in the case of pure
fermions. Thus, the  many-particle system under study is effectively
described (and interpreted) in terms of the deformed one, i.e., by
using (quasi)particles with a non-standard statistics.

\section{Deformed Bose Gas which Accounts\\ for the Compositeness of
Particles} \label{sec:compos}

In this section, we consider a $\tilde{\mu}$-deformed Bose gas with
the quadratic structure function
\begin{equation}
\phi_{\tilde{\mu}}(n)\equiv [n]_{\tilde{\mu}} = (1+\tilde{\mu})n -
\tilde{\mu} n^2,
\end{equation}
which was shown to realize (under certain conditions,
see~\cite{GKM,GKM2}) the gas of two-fermion composite Bose-like
particles. For such a deformation, we now derive the deformed virial
expansion of the equation of state along with a few first virial
coefficients, by using the concept of $\tilde{\mu}$-deformed
derivative.

As a starting point, we take the logarithm of the Bose gas grand
partition function:
\begin{equation}
\ln Z = - \sum_i \ln (1-z e^{-\beta \varepsilon_i}).
\end{equation}
Instead of the Jackson derivative $\mathcal{D}^{q}$ used in
\cite{Scarfone_Interact} or its $p,q$-extension
$\mathcal{D}^{(p,q)}$ exploited in \cite{GR_Virial}, we apply the
following $\tilde{\mu}$-deformed extension of $\frac{d}{dz}$ defined
through its action on monomials $z^k$:
\[\mathcal{D}_z^{(\tilde{\mu})} z^k = ((1+\tilde{\mu})k - \tilde{\mu}
k^2) z^{k-1}.\] This can also be obtained by the action of the
operator
\[
z \mathcal{D}_z^{(\tilde{\mu})} =
\Bigl[z\frac{d}{dz}\Bigr]_{\tilde{\mu}} = \Bigl((1+\tilde{\mu})
\Bigl(\!z\frac{d}{dz}\Bigr) - \tilde{\mu}
\Bigl(\!z\frac{d}{dz}\Bigr)^2\Bigr)
\]
on monomials. Moreover, the latter acts in an obvious way on an
arbitrary smooth function $f(z)=$ $=\sum_{j=0}^\infty c_j z^j$.

For the grand partition function logarithm, we have the
following expansion in~$z$:
\begin{align*}
&\ln Z = \frac{\pi\sqrt{\pi}V}{(2\pi)^3}
\Bigl(\frac{2m}{\beta\hbar^2}\Bigr)^{\!3/2} \sum_{n=1}^\infty
\frac{z^n}{n^{5/2}}=\\
&=  \frac{V}{\lambda^3} \sum_{n=1}^\infty \frac{z^n}{n^{5/2}}
\end{align*}
where $\lambda = h/(2\pi m k_{\rm  B} T)^{1/2}$ is the thermal
wavelength. Then, in case of the deformed picture, we obtain that
the number of particles $N^{(\tilde{\mu})}$ according to the above
consideration is
\begin{equation}
N^{(\tilde{\mu})} \!=\! \Bigl[z\frac{d}{dz}\Bigr]_{\tilde{\mu}}\!
\ln Z \!\equiv\! z \mathcal{D}_z^{(\tilde{\mu})}\! \ln Z \!=\!
\frac{V}{\lambda^3}\! \sum_{n=1}^\infty [n]_{\tilde{\mu}}
\frac{z^n}{n^{5/2}}\label{N^mu}
\end{equation}
or, equivalently,
\begin{equation}\label{lambda/v}
\frac{\lambda^3}{v} = \sum_{n=1}^\infty [n]_{\tilde{\mu}} \frac{z^n}{n^{5/2}},
\end{equation}
where the specific volume $v=\frac{V}{N^{(\tilde{\mu})}}$ is
introduced. The equation of state for a quantum gas of
non-interacting, but composite particles
\begin{equation}
\frac{PV}{k_{\rm  B} T} = \ln Z = \frac{V}{\lambda^3}
\Bigl(z+\frac{z^2}{2^{5/2}}+\frac{z^3}{3^{5/2}}+\frac{z^4}{4^{5/2}}+\frac{z^5}{5^{5/2}}+...\Bigr)
\end{equation}
is now deformed in the following way:\vspace*{-1mm}
\begin{equation}
\frac{PV}{k_{\rm  B} T} = \ln Z^{(\tilde{\mu})}.
\end{equation}
The  desired $\tilde{\mu}$-deformed partition function
$Z^{(\tilde{\mu})}$ is defined from the deformed analog of the
relation $N=\bigl(z\frac{d}{dz}\bigr) \ln Z$, i.e., from $
N^{(\tilde{\mu})} = \Bigl(z\frac{d}{dz}\Bigr)\ln Z^{(\tilde{\mu})}$
or\vspace*{-3mm}
\begin{equation}
\ln Z^{(\tilde{\mu})} = \Bigl(z\frac{d}{dz}\Bigr)^{\!-1}
N^{(\tilde{\mu})}.
\end{equation}
As a result, we obtain the following expansion for the deformed
equation of state:\vspace*{-1mm}
\begin{multline}\label{eq_st(z)}
\frac{PV}{k_{\rm  B} T} = \ln Z^{(\tilde{\mu})} =\\
= \frac{V}{\lambda^3}
\Bigl(\!z+\frac{[2]_{\tilde{\mu}}}{2^{7/2}}z^2+\frac{[3]_{\tilde{\mu}}}{3^{7/2}}z^3+
\frac{[4]_{\tilde{\mu}}}{4^{7/2}}z^4+\frac{[5]_{\tilde{\mu}}}{5^{7/2}}z^5+...\Bigr).
\end{multline}
In order to deduce the corresponding virial expansion, we have to
find the function $z(\lambda^3/v)$ in the form of a series, through
inverting (\ref{lambda/v}). For this purpose, we expand
$z(\lambda^3/v) = z(\lambda^3 N/V)$ in a Taylor series
as\vspace*{-1mm}
\[
z(\lambda^3/v) = z'_N|_{N=0} N + \frac{z''_N}{2!}\Bigr|_{N=0} N^2 +
\frac{z'''_N}{3!}\Bigr|_{N=0} N^3 +
\]\vspace*{-7mm}
\begin{equation}\label{z(N)}
+ \frac{z^{(IV)}_N}{4!}\Bigr|_{N=0} N^4 +
\frac{z^{(V)}_N}{5!}\Bigr|_{N=0} N^5 +...\,.
\end{equation}
The derivatives $z^{(l)}_N|_{N=0}$, $l=1,2,...$, can be expressed
through analogous derivatives $N^{(r)}_z|_{z=0}$, $r=1,2,...$\,.
>From (\ref{N^mu}), we find the $r$-th order derivative of
$N^{(\tilde{\mu})}$ with respect to $z$ at $z=0$:\vspace*{-1mm}
\begin{equation}
N^{(r)}_z\Bigr|_{z=0} = \frac{\pi\sqrt{\pi}V}{(2\pi)^3}
\Bigl(\frac{2m}{\beta\hbar^2}\Bigr)^{3/2} r!
\frac{[r]_{\tilde{\mu}}}{r^{5/2}} = \frac{V}{\lambda^3} r!
\frac{[r]_{\tilde{\mu}}}{r^{5/2}}.\!\!\!
\end{equation}
For $z'_N|_{N=0}$, ...,$z^{(V)}_N|_{N=0},$ we infer\vspace*{-1mm}
\begin{align*}
&z'_N|_{N=0} = \frac{1}{N'_z}\Bigr|_{N=0} = \frac{\lambda^3}{V} \frac{1}{[1]_{\tilde{\mu}}},\\
&z''_N|_{N=0} = -\frac{N''_z}{(N'_z)^3} = -\Bigl(\frac{\lambda^3}{V}\Bigr)^2 \frac{2!}{2^{5/2}}
\frac{[2]_{\tilde{\mu}}}{[1]_{\tilde{\mu}}^3},\\
&z'''_N|_{N=0} = \Bigl(-\frac{N'''_z}{(N'_z)^5}+3\frac{(N''_z)^2}{(N'_z)^5}\Bigr)\Bigr|_{z=0}
=\\
&= \Bigl(\frac{\lambda^3}{V}\Bigr) \Bigl(3 \frac{(2!)^2}{2^5}
\frac{[2]_{\tilde{\mu}}}{[1]^5_{\tilde{\mu}}} - \frac{3!}{3^{5/2}}
\frac{[3]_{\tilde{\mu}}}{[1]^4_{\tilde{\mu}}}\Bigr),\\
&...\,.
\end{align*}
Substituting these derivatives into~(\ref{z(N)}), we find
\[
z(\lambda^3/v) =\frac{1}{[1]_{\tilde{\mu}}} \frac{\lambda^3}{v} -
\frac{1}{2^{5/2}} \frac{[2]_{\tilde{\mu}}}{[1]^3_{\tilde{\mu}}}
\Bigl(\frac{\lambda^3}{v}\Bigr)^{\!2} +\]\vspace*{-7mm}
\[+ \Bigl(\frac{1}{2^4}
\frac{[2]^3_{\tilde{\mu}}}{[1]^5_{\tilde{\mu}}} - \frac{1}{3^{5/2}}
\frac{[3]_{\tilde{\mu}}}{[1]^4_{\tilde{\mu}}}\Bigr)
\Bigl(\frac{\lambda^3}{v}\Bigr)^{\!3} +
\]\vspace*{-7mm}
\[
+ \Bigl(-\frac{1}{4^{5/2}}
\frac{[4]_{\tilde{\mu}}}{[1]^5_{\tilde{\mu}}} +
\frac{5}{2^{5/2}3^{5/2}}
\frac{[2]_{\tilde{\mu}}[3]_{\tilde{\mu}}}{[1]^6_{\tilde{\mu}}} -
\frac{5}{2^{15/2}}
\frac{[2]^3_{\tilde{\mu}}}{[1]^7_{\tilde{\mu}}}\Bigr)
\Bigl(\!\frac{\lambda^3}{v}\!\Bigr)^{\!4} +
\]\vspace*{-7mm}
\[
+ \Bigl(-\frac{1}{5^{5/2}}
\frac{[5]_{\tilde{\mu}}}{[1]^6_{\tilde{\mu}}} + \frac{3}{2^{13/2}}
\frac{[2]_{\tilde{\mu}}[4]_{\tilde{\mu}}}{[1]^7_{\tilde{\mu}}} +
\frac{1}{3^4} \frac{[3]^2_{\tilde{\mu}}}{[1]^7_{\tilde{\mu}}} -
\]\vspace*{-6mm}
\begin{equation}
- \frac{7}{2^5 3^{3/2}}
\frac{[2]^2_{\tilde{\mu}}[3]_{\tilde{\mu}}}{[1]^8_{\tilde{\mu}}} +
\frac{7}{2^9} \frac{[2]^4_{\tilde{\mu}}}{[1]^9_{\tilde{\mu}}}\Bigr)
\Bigl(\frac{\lambda^3}{v}\Bigr)^{\!5} + ...\,.
\end{equation}
Plugging this expression into~(\ref{eq_st(z)}) we arrive at the
desired virial expansion depending on the parameter $\tilde{\mu}$,
which corresponds to a {\it non-interacting gas of composite
bosons}:\vspace*{-1mm}
\[
\frac{P}{k_{\rm  B} T} = v^{-1} \biggl\{\sum_{k=1}^\infty
V_k(\tilde{\mu}) \Bigl(\frac{\lambda^3}{v}\Bigr)^{k-1}\biggr\} =
\]\vspace*{-7mm}
\[
= v^{-1} \Bigl\{1 - \frac{[2]_{\tilde{\mu}}}{2^{7/2}}
\frac{\lambda^3}{v} + \Bigl(\frac{[2]^2_{\tilde{\mu}}}{2^5} -
\frac{2[3]_{\tilde{\mu}}}{3^{7/2}}\Bigr)\Bigl(\frac{\lambda^3}{v}\Bigr)^{\!2}
+
\]\vspace*{-7mm}
\[
+ \Bigl(-\frac{3[4]_{\tilde{\mu}}}{4^{7/2}} +
\frac{[2]_{\tilde{\mu}}[3]_{\tilde{\mu}}}{2^{5/2}3^{3/2}} -
\frac{5[2]^3_{\tilde{\mu}}}{2^{17/2}}\Bigr)\Bigl(\frac{\lambda^3}{v}\Bigr)^{\!3}
+
\]\vspace*{-7mm}
\[
+ \Bigl(-\frac{4[5]_{\tilde{\mu}}}{5^{7/2}} +
\frac{[2]_{\tilde{\mu}}[4]_{\tilde{\mu}}}{2^{11/2}} -
\frac{2[3]^3_{\tilde{\mu}}}{3^5} -
\frac{[2]^2_{\tilde{\mu}}[3]_{\tilde{\mu}}}{2^3 3^{3/2}} +
\frac{7[2]^4_{\tilde{\mu}}}{2^{10}}\Bigr)\times
\]\vspace*{-7mm}
\begin{equation}\label{Virial_mu}
 \times
\Bigl(\frac{\lambda^3}{v}\Bigr)^{\!4} + ...\Bigr\}.
\end{equation}
Here, $V_k(\tilde{\mu})$, $k=1,2,...$, are the virial coefficients.
Note that, unlike (5) and [23], the obtained virial coefficients are
found exactly.

Let us make the observation that the second virial coefficient $V_2
= - \frac{[2]_{\tilde{\mu}}}{2^{7/2}} = -
\frac{1-\tilde{\mu}}{2^{5/2}}$ vanishes, when $\tilde{\mu}$ reaches
$1$. This can be interpreted as a compensation, at $\tilde{\mu}=1$,
of the compositeness effects, against the quantum-statistical
many-particle effects, so that the quantum gas of composite bosons
behaves itself like a classical gas of pointlike particles, at least
to $\lambda^3/v$-terms. An analogous statement may be deduced also
for the next virial coefficients $V_3$, $V_4$, ...\,.

Like before, the modified virial coefficients, in the absence of the
explicit interaction between composite bosons in the Hamiltonian,
can be viewed, in the current interpretation, as those reflecting an
effective interaction of the quantum statistical origin between
bosons (like an effective attraction (respectively, a repulsion) of
the quantum origin between bosons (respectively, fermions)).

\section{Account for Both Compositeness\\ and Interaction of Particles}

To consider the interaction jointly with the compositeness of the
particles of a gas, we use the function\vspace*{-4mm}
\begin{equation}\label{phi_mu_q}
\phi_{\tilde{\mu},q}(n) = (1+\tilde{\mu})[n]_q - \tilde{\mu}
([n]_q)^2
\end{equation}\vspace*{-5mm}

\noindent and the respective $(\tilde{\mu},q)$-extension of the
Jackson derivative\vspace*{-2mm}
\[
z \mathcal{D}_z^{(\tilde{\mu},q)} = \Bigl((1+\tilde{\mu})
\Bigl[z\frac{d}{dz}\Bigr]_{q} - \tilde{\mu}
\Bigl[z\frac{d}{dz}\Bigr]_{q}^2\Bigr).
\]\vspace*{-4mm}

\noindent This latter acts in an obvious way on monomials and on an
arbitrary smooth function $f(z)=\sum_{j=0}^\infty c_j z^j$.

We have to remark that, similarly to  
the Hamiltonian $H_{\epsilon}$ in (4), the two-parameter Hamiltonian
$H_{\tilde{\mu},\epsilon}=\frac12\omega\bigl([N+1]_{\tilde{\mu},q}+[N]_{\tilde{\mu},q}\bigr)$
for the system of ${\tilde{\mu},q}$-bosons (single-mode case) can be
split, in the present case, into the Hamiltonian $H_0$ (with no
interaction and no compositeness) and the double-interaction
Hamiltonian $H_1(\epsilon,\tilde{\mu}; N)$ that depends on $N$ and
is the double series in $\tilde{\mu}$ and $\epsilon=q-1$. The
compositeness ``causes'' the extra amount of an effective
interaction of the quantum origin, on the equal footing with that
encoded in $q=1+\epsilon$.

Proceeding along lines similar to the previous sections, we can
obtain the virial expansion in the case under consideration, using
the deformation structure function $\phi_{\tilde{\mu},q}(n)$ and the
deformed derivative $\mathcal{D}_z^{(\tilde{\mu},q)}$. We note that
$\mathcal{D}_z^{(\tilde{\mu})}$ and
$\mathcal{D}_z^{(\tilde{\mu},q)}$ have similar definitions in terms
of the respective structure functions, and
$[1]_{\tilde{\mu}}=[1]_{\tilde{\mu},q}=1.$ So, we infer the desired
formulas by making the replacement $[k]_{\tilde{\mu}}\rightarrow
[k]_{\tilde{\mu},q}$, $k=2,3,4,5,...$ in (\ref{Virial_mu}). The
resulting virial expansion of the equation of state takes the
form\vspace*{-1mm}
\[
\frac{Pv}{k_{\rm  B} T} =\biggl\{\sum_{k=1}^\infty
V_k(\tilde{\mu},q) \Bigl(\frac{\lambda^3}{v}\Bigr)^{k-1}\biggr\}=
\Bigl\{1 - \frac{[2]_{\tilde{\mu},q}}{2^{7/2}} \frac{\lambda^3}{v} +
\]\vspace*{-4mm}
\[
+\Bigl(\frac{[2]^2_{\tilde{\mu},q}}{2^5} -
\frac{2[3]_{\tilde{\mu},q}}{3^{7/2}}\Bigr)\Bigl(\frac{\lambda^3}{v}\Bigr)^{\!2}
+ \Bigl(-\frac{3[4]_{\tilde{\mu},q}}{4^{7/2}} +
\frac{[2]_{\tilde{\mu},q}[3]_{\tilde{\mu},q}}{2^{5/2}3^{3/2}}-
\]\vspace*{-4mm}
\[
-
\frac{5[2]^3_{\tilde{\mu},q}}{2^{17/2}}\Bigr)\Bigl(\frac{\lambda^3}{v}\Bigr)^{\!3}
+ \Bigl(-\frac{4[5]_{\tilde{\mu},q}}{5^{7/2}} +
\frac{[2]_{\tilde{\mu},q}[4]_{\tilde{\mu},q}}{2^{11/2}}-
\]\vspace*{-4mm}
\begin{equation}\label{Virial_mu_q}
- \frac{2[3]^3_{\tilde{\mu},q}}{3^5} -
\frac{[2]^2_{\tilde{\mu},q}[3]_{\tilde{\mu},q}}{2^3 3^{3/2}}+
\frac{7[2]^4_{\tilde{\mu},q}}{2^{10}}\Bigr)\Bigl(\frac{\lambda^3}{v}\Bigr)^{\!4}
+ ...\Bigr\}.
\end{equation}

It is now natural to interpret the virial expansion
(\ref{Virial_mu_q}) as the effective one corresponding to the
interacting gas of composite (quasi)particles. The information about
the interaction and the composite structure is merely encoded in the
two deformation parameters $q$ and $\tilde{\mu},$ respectively. In
the limiting case $\tilde{\mu}=0$, expansion~(\ref{Virial_mu_q})
accounts only for the interaction between the particles; likewise,
when $q=1,$ formula (\ref{Virial_mu_q}) should be interpreted as
accounting only for the compositeness of particles. When $q\ne1$ and
$\tilde{\mu}\ne0$, expression (\ref{Virial_mu_q}) incorporates the
both mentioned factors of non-ideality of a Bose gas jointly.

Consider a deviation of the second virial coefficient $V_2$ from its
non-deformed value in the limiting cases where $\tilde{\mu}=0$ or
respectively $q=1$, namely,
\begin{equation}\label{Delta_V_2}
\Delta V_2 \mathop{\longrightarrow}\limits^{\tilde{\mu}=0} \frac{1-q}{2^{7/2}}\quad {\rm resp.}\quad
\Delta V_2 \mathop{\longrightarrow}\limits^{q=1} \frac{\tilde{\mu}}{2^{5/2}}.
\end{equation}
Remark that, for these deviations, the explicit formulas through
interaction potentials could be obtained (this is, however, a rather
non-trivial problem being beyond the scope of this paper) from the
corresponding expressions for the virial coefficients (see, e.g.,
\cite{Landau_5,Huang,Pathria}). By comparing those explicit formulas
with~(\ref{Delta_V_2}), it is possible to relate the deformation
parameters $q$ and $\tilde{\mu}$ with the parameters in the
Hamiltonian of interaction between composite bosons and inside them
(i.e., between the constituents).

\section{An Alternative Description\\ Using Other Structure Functions}

The deformation structure function
\mbox{$\phi_{\tilde{\mu},q}(N)\equiv $}\linebreak
$\equiv\phi_{\tilde{\mu}}(\phi_q(N))$ from (\ref{phi_mu_q}) is not
the only possible one. Indeed, some other structure functions can be
used to take the compositeness along with the interaction into account. For
instance, the function
\begin{multline}\label{phi_q_mu}
\phi_{q,\tilde{\mu}}(N)\equiv \phi_q(\phi_{\tilde{\mu}}(N)) \equiv
 \frac{1\!-\!q^{[N]_{\tilde{\mu}}}}{1\!-\!q} \equiv \frac{1\!-\!q^{(1\!+\!\tilde{\mu})N\!-\!\tilde{\mu}N^2}}{1\!-\!q}
\end{multline}
also possesses the limiting cases
\mbox{$\phi_{q,\tilde{\mu}}|_{q=1}(N)=$}\linebreak
$=\phi_{\tilde{\mu}}(N)$,
$\phi_{q,\tilde{\mu}}|_{\tilde{\mu}=0}(N)=\phi_q(N).$ Thus, it can
be used as another admissible structure function instead of
$\phi_{\tilde{\mu},q}(N)$. Moreover, from the structure functions
$\phi_{\tilde{\mu},q}(N)$ and $\phi_{q,\tilde{\mu}}(N),$ we can form
the whole one-parameter family of structure functions
\begin{equation}\label{phi_t}
\phi_t(N) \equiv\phi_{t;\tilde{\mu},q}(N)= t \phi_{\tilde{\mu},q}(N)
+ (1-t) \phi_{q,\tilde{\mu}},
\end{equation}
which are related to the above-mentioned one-parameter limits (20).
The corresponding virial coefficients, by exploiting the analogy
to~(\ref{Virial_mu_q}), are the following:\vspace*{-3mm}
\begin{equation}
\begin{array}{l}
\displaystyle V_1= 1,\ V_2(t; \tilde{\mu}, q)  =  -  \frac{\phi_{t
;\tilde{\mu},q}(2)}{2^{7/2}},\\[2mm]
\displaystyle V_3(t; \tilde{\mu}, q) = \frac{(\phi_{t ;\tilde{\mu},q}(2))^2}{2^5}  -  \frac{2\phi_{t ;\tilde{\mu},q}(3)}{3^{7/2}},\\
\displaystyle V_4(t; \tilde{\mu}, q) =\\ \displaystyle -
\frac{3\phi_{t ;\tilde{\mu},q}(4)}{4^{7/2}}  + \frac{\phi_{t
;\tilde{\mu},q}(2)\phi_{t ;\tilde{\mu},q}(3)}{2^{5/2}3^{3/2}}
 -  \frac{5(\phi_{t ;\tilde{\mu},q}(2))^3}{2^{17/2}},\\[2mm]
\displaystyle V_5(t; \tilde{\mu}, q) =\\
\displaystyle - \frac{4\phi_{t ;\tilde{\mu},q}(5)}{5^{7/2}}  +
\frac{\phi_{t ;\tilde{\mu},q}(2)\phi_{t
;\tilde{\mu},q}(4)}{2^{11/2}}
 -  \frac{2(\phi_{t ;\tilde{\mu},q}(3))^3}{3^5} - \\[2mm]
\displaystyle - \frac{(\phi_{t ;\tilde{\mu},q}(2))^2\phi_{t
;\tilde{\mu},q}(3)}{2^3 3^{3/2}} + \frac{7(\phi_{t
;\tilde{\mu},q}(2))^4}{2^{10}}.
\end{array}\label{24}\!\!\!\!\!\!\!\!
\end{equation}
Of course, the further study on more physical (phenomenological)
grounds is needed in order that the preference could be made for one
deformed model, see~(\ref{Virial_mu_q}), with respect to those
contained in (22).\vspace*{-2mm}

\section{Conclusions}

Basing on~\cite{GKM,GKM2} and \cite{Scarfone_Interact}, the
specially designed two-parameter deformed Bose gas model capable to
(effectively) describe the interacting gas of composite bosons is
constructed. The specific deformation structure function, which
characterizes the deformed bosons (oscillators) of this model, is
constructed by combining the previously studied ones
from~\cite{GKM,GKM2} and \cite{Scarfone_Interact}. The ``building
block'' structure functions are the quadratic polynomial structure
function of deformation and the $q$-deformed structure function of
the Arik--Coon type which provides the effective description of an
interacting gas of elementary bosons.

For the proposed deformed Bose gas model, the corresponding deformed
virial expansion is found along with the first five virial
coefficients and interpreted as the virial expansion accounting for
both the interaction of composite bosons and their composite
structure. The thermodynamic relations for the deformed Bose gas
model (including the equation of state), which were utilized in the
process of derivation of its virial expansion, are obtained, by
using an appropriate generalization of the Jackson derivative for
the concerned unified deformation.

Some alternative deformation structure functions for the deformed
Bose gas model needed for the effective description are proposed.
Those include the composition of the $q$-deformed structure function
and the quadratic one, taken in the opposite order,
see~(\ref{phi_q_mu}), and the $t$-parameter family interpolating
between the both, see~(\ref{phi_t})--(\ref{24}).

Note that it is also of interest to  examine the correlation
function intercepts for the considered deformed Bose gas model in
conjunction with the above-mentioned interpretation and to make
comparison with experimental data (e.g., those on the
$\pi^-$-mesons, the known composites). For some other deformed Bose
gas models, the comparison of $\pi$-mesonic correlation functions
intercepts of the second (and third) order with experimental data is
already done in a few previous papers, see, e.g.,
\cite{Anchishkin_Transverse,Gavrilik_Sigma,GR_EPJA}.

\vskip3mm {\it This  work was partly supported by the Special
Program of the Division of Physics and Astronomy of the NAS of
Ukraine, and by the Grant (A.P.R.) for Young Scientists of the NAS
of Ukraine No.~0113U004910.}

%


\vspace*{-5mm} \rezume{О.М.\,Гаврилик, Ю.А.\,Міщенко}{МОДЕЛІ
ДЕФОРМОВАНОГО БОЗЕ-ГАЗУ,\\ ЗДАТНІ ОДНОЧАСНО ВРАХУВАТИ СКЛАДЕНУ\\
СТРУКТУРУ ЧАСТИНОК ТА ЇХ ВЗАЄМОДІЮ}  {В даній роботі ми розглядаємо
модель деформованого бозе-газу зі структурною функцією, яка є
комбінацією $q$-деформації та квадратичної поліноміальної
деформації. Такий вибір уніфікованої структурної функції деформації
дає можливість описати взаємодіючий газ складених (двоферміонних чи
двобозонних) бозонів. Використовуючи відповідне узагальнення
похідної Джексона, отримано деформований віріальний розклад.
Останній інтерпретується як віріальний розклад для газу взаємодіючих
складених бозонів із деяким потенціалом взаємодії.}

\end{document}